\documentclass[twocolumn,showpacs,preprintnumbers,amsmath,amssymb]{revtex4}
\usepackage{graphicx}% Include figure files
\usepackage{dcolumn}% Align table columns on decimal point
\usepackage{bm}% bold math

\begin{document}

\title{The Low-Frequency Character of the Thermal Correction to the Casimir Force between
Metallic Films\\
LA-UR 02-4696}

\author{J.R. Torgerson and S.K. Lamoreaux}

\affiliation{University of California,Los Alamos National
Laboratory,Physics Division P-23, M.S. H803, Los Alamos, NM 87545}

\date{today}

\begin{abstract}

The frequency spectrum of the finite temperature correction to the
Casimir force can be determined by use of the Lifshitz formalism
for metallic plates of finite conductivity. We show that the
correction for the $TE$ electromagnetic modes is dominated by
frequencies so low that the plates cannot be modelled as ideal
dielectrics.  We also address issues relating to the behavior of
electromagnetic fields at the surfaces and within metallic
conductors, and calculate the surface modes using appropriate
low-frequency metallic boundary conditions. Our result brings the
thermal correction into agreement with experimental results that
were previously obtained.

\end{abstract}
\pacs{12. 20. Ds, 41. 20. -q, 12. 20. Fv}
\maketitle

\section{Introduction}

A recent paper \cite{1}, in which simultaneous consideration of
the thermal and finite conductivity corrections to the Casimir
force between metal plates leads to a significant deviation from
an experimental result \cite{2} and previous theoretical work, has
attracted considerable interest. The principal conclusion in
\cite{1} leading to this discrepancy is that the $TE$
electromagnetic mode (${\bf E}$ parallel to the surface) does not
contribute to the force at finite temperature. Arguments against
the analysis given in \cite{1} have been numerous \cite{3,4,5,6}
but not universally accepted \cite{7,8}.

A careful numerical analysis of the problem leads us and others to
conclude that the results presented in \cite{1} are mathematically
correct.  As we show here, this analysis does not accurately
represent the experimental arrangement used in \cite{2}. The
aspect of the problem that has not been considered in detail is
the appropriateness of a dielectric model of the metallic plates
at low frequencies, which, as we will show, are most relevant for
the thermal correction. The purpose of this note is to expand on
our previous work \cite{9} and to point out that the proper
boundary conditions for conductors have not yet been directly
applied to this problem, and to show that the experimental result
\cite{2} can be fully explained by this application.

\section{Spectrum of the $TE$ Mode Thermal Correction of the Casimir Force}

Following Ford \cite{10}, the spectrum of the Casimir force is
given by Eqs. (2.3) and (2.4) of Lifshitz' seminal paper
\cite{11}. We note that
\begin{equation}
{1\over 2}\coth {\hbar\omega\over 2kT}={1\over 2} + {1\over
\exp(\hbar\omega/ kT)-1}={1\over 2} + g(\omega)
\end{equation}
and we only include the second term on the right-hand side in the
determination of the spectrum of the thermal correction.  From Eq.
(2.4) of \cite{11}, the spectrum of the $TE$ mode excitation
between parallel plates can be described by
\begin{eqnarray}\label{lif}
\left[{\hbar\over \pi^2 c^3}\right]F_\omega&=&\left[{\hbar\over
\pi^2 c^3}\right]\omega^3 g(\omega)\cr &\times &{\rm Re}\int_C
p^2dp
 \left[{(s+p)^2\over(s-p)^2}e^{-2ip\omega a/c}-1\right]^{-1},
\end{eqnarray}
\begin{equation}\label{s}
s=\sqrt{\epsilon(\omega)-1+p^2}
\end{equation}
where $a$ is the plate separation, and we have assumed that the
plates are made of the same material with vacuum between them. The
integration path $C$ can be separated into $C_1$ for  $p=1$ to 0,
which describes the effect of plane waves, and $C_2$ with pure
imaginary values $p=i0$ to $i\infty$ for exponentially damped
(evanescent) waves.

In anticipation that the effect is a low-frequency phenomenon, we
use the parameters for Au in \cite{1} for ${\rm Im}\
\epsilon=\epsilon_2$  and employ the Kramers-Kronig relations to
determine ${\rm Re}\ \epsilon=\epsilon_1$. We find for frequencies
$\omega<10^{14}$ s$^{-1}$ that, to good approximation,
\begin{equation}\label{epsilon}
\epsilon_1={-1.48\times 10^{4}\over 1+(\omega/\omega_0)^2};\ \
\epsilon_2={1.8\times 10^{18}\over \omega(1+(\omega/\omega_0)^2)}
\end{equation}
with $\omega_0=3.3\times 10^{13}\ {\rm s}^{-1}$.

In \cite{1}, a net deviation from the zero-temperature value of
the Casimir force is predicted to be about 25\% for a plate
separation of $1\ \mu$m at 300 K.  The experimental results
reported in \cite{2} had their greatest sensitivity around 1
$\mu$m, and disagree significantly with the results in \cite{1}.
As a comparison, we numerically integrate Eq. {\ref{lif}) for
$a=1\ \mu$m and $T=300$ K, using Eq. (\ref{epsilon}) for the
permittivity. The results are shown in Fig. 1, where we have
separated the results from the two integration paths.   In Fig.
1a, it can be seen that there is no significant deviation from the
perfectly conducting case. On the other hand, the contribution
from evanescent waves, shown in Fig. 1b, is large and the
integrated value is in good agreement with the result given in
\cite{1}.

We see immediately that the main contributions of the $TE$-mode
finite conductivity correction are around $\omega=10^{10}-10^{13}$
s$^{-1}$.  This behavior is due to an approximately quadratic
increase with $\omega$ of the $C_2$ integral and a suppression
beginning at $\omega=kT/\hbar=4\times 10^{13}$ s$^{-1}$ due to
$g(\omega)$. This is a low frequency range and we can question
certain assumptions in \cite{1} and in the Lifshitz calculation,
among others, in regard to theoretical predictions relevant to the
experimental arrangement in \cite{2}.

\section{Low Frequency Limit and Field Behavior in Metallic Materials}

When the depth of penetration of the electromagnetic field into a
metal,
\begin{equation}
\delta=c/\sqrt{2\pi\mu\sigma\omega}
\end{equation}
where $\sigma$ is the conductivity and $\mu$ is the permeability
(for Au and Cu, $\sigma\approx 3\times 10^{17}\ {\rm s}^{-1}$,
$\mu=1$), becomes of the same order as the mean free path of the
conduction electrons, it is no longer possible to describe the
field in terms of a dielectric permeability \cite{12,13}. This
occurs for optical frequencies $\omega\approx 5\times 10^{13}$
s$^{-1}$ for metals such as Au and Cu where the mean free path, at
300 K, is about $3\times 10^{-6}$ cm \cite{14} (p. 259). At
frequencies above $10^{14}$ s$^{-1}$ the permeability description
again becomes valid because on absorbing a photon, a conduction
electron acquires a large kinetic energy and has a shortened mean
free path. However, in the interaction of a field with a material
surface, $\bf E$ and $\bf H$ can always be related linearly
through the surface impedance (which relates the electric field at
the surface to a surface current hence magnetic field); this
approach has been used in calculation of the Casimir force
\cite{15}. A related correction arises from the the plasmon
interaction with the surface which becomes significant near the
plasma frequency of the metal, and has been estimated as nearly
10\%\space \cite{16} for sub-$\mu$m plate separations.

The proper boundary conditions for a conducting plane have been
discussed by Boyer \cite{17}. He points out that when (using here
the notation of \cite{1}) $\omega\ll\eta^2\rho/4\pi$, where $\rho$
is the resistivity and $\eta$ is the dissipation, the usual
dielectric boundary conditions are not applicable. For Au, using
the parameters in \cite{1}, this limit is met for
$\omega\ll4\times 10^{14}\ {\rm s}^{-1}$. This corresponds to an
optical wavelength of 5\,$\mu$m, which implies that for plate
separations significantly larger than this, and of course for
$\omega\rightarrow 0$, the plates must be treated as good
conductors.

The boundary conditions for a conducting surface are discussed in
\cite{18} (Sec. 8.1). At low frequencies (e.g., where the
displacement current can be neglected), a tangential electric
field at the surface of a conductor will induce a current ${\bf
j}_\|=\sigma {\bf E}_\|$, where $\sigma$ is the conductivity. The
presence of the surface current leads to a discontinuity in the
normal derivative of ${\bf H}_\|$, hence a discontinuity in the
normal derivative of ${\bf E}_\|$, at the boundary of a conducting
surface.  These boundary conditions are quite different from the
dielectric case where the fields and their derivatives are assumed
continuous.

\section{Electromagnetic Modes between Metallic Plates}

We are interested in modes between two conducting plates separated
by a distance $a$. In the limit that the plates are thin films of
thickness $d>\delta$, the skin depth, we can assume that the
plates are infinitely thick and the problem is considerably
simplified. This is well-satisfied for the conditions of the
experiment \cite{2} when $\omega>10^{11}\ {\rm s^{-1}}$ in which
case $\delta< 0.7\ \mu$m compared to the film thickness of 1
$\mu$m. Essentially all of the $TE$ mode thermal correction comes
in the $10^{11}$ and $10^{13}$ s$^{-1}$ range as shown in Fig. 1.

Taking the $\hat z$ axis as perpendicular to the plates, and the
mode propagation direction along $\hat x$, for the case of $TE$
modes (also referred to as $H$ or magnetic modes), $E_x=0$. The
plates surfaces are located at $z=0$ and $z=a$.  For a perfect
conductor, $\partial H_z/\partial z=0$ at the conducting surfaces.
A finite conductivity makes this derivative non-zero, and can be
estimated from the small electric field $E_y$ that exists at the
surface of the plate, (see \cite{18}, Sec.  8.1 and Eq. (8.6))
\begin{equation}
\vec E_\|=\hat y E_y=\sqrt{\omega\over 8\pi \sigma}(1-i)\hat
n\times\vec H_\|.
\end{equation}
where $\vec H\|= \hat x H_x$ and it is assumed that the
displacement current in the metal plate can be neglected
($\sigma>>\omega$), and that the inverse of the mode wavenumber is
less than $\delta$. $E_y$ and $H_x$ are related through Maxwell's
equation $\vec\nabla\times\vec H=\partial \vec E/c\partial t$.
Assuming a time dependence of $e^{-i\omega t}$, and vacuum between
the plates,
\begin{equation}\label{diffeq}
{\partial H_x\over \partial z}=\pm {i\omega\over c} E_y.
\end{equation}
where $\pm$ indicates sign of $\hat n$ at $z=0$ and $z=a$
respectively.  The boundary conditions at the surfaces are thus
\begin{equation}
{\partial\over\partial z}H_z=\pm i\sqrt{\omega\over 8 \pi
\sigma}\left({\omega\over c}\right) (1-i)H_z\equiv \pm \alpha H_z.
\end{equation}
Solutions of the form $H_\|(z)=Ae^{Kz}+Be^{-Kz}$, where
$K^2=k^2-\omega^2/c^2$ and $k$ is the transverse wavenumber, can
be constructed for the space between the conducting plates. The
eigenvalues $K$ can be determined by the requirement that Eq.
(\ref{diffeq}) be satisfied at $z=0$ and $z=a$. With the usual
substitution $\omega=i\xi$, the eigenvalues $K$ are then given by
(see \cite{19}, Sec. 7.2)
\begin{equation}\label{gte}
G_{TE}(\xi)\equiv{(\alpha + K)^2\over (\alpha-K)^2}e^{2Ka}-1=0
\end{equation}
and the force can be calculated by the techniques outlined in
\cite{19}, Sec. 7.3.

This result can be recast in the notation of the Lifshitz
formalism, and the spectrum of the thermal correction can be
calculated as before. Noting that $K=i\omega p/c$,
\begin{equation}
F_\omega=\omega^3g(\omega)\int_C p^2dp\left[{(\alpha+i\omega
p/c)^2\over(\alpha-i\omega p/c)^2}e^{-2i\omega p/c}-1\right]^{-1}.
\end{equation}
Results of a numerical integration are shown in Fig. 2, where it
can be seen by comparison with Fig. 1 that the metallic plate
boundary condition does not show a significant contribution from
the $C_2$ integral of the $TE$ mode thermal correction and is
therefore similar to that for the ``perfect conductor'' boundary
condition. This reconciles the discrepancy between the prediction
in \cite {1} and the experimental results reported in \cite{2}.

\section{Conclusion}

The problem of calculating the $TE$ mode contribution to the
Casimir force has been previously treated with the  ``Schwinger
prescription'' \cite{20} of setting the dielectric constant to
infinity before setting $\omega=0$. This prescription has become
controversial \cite{21}, a term that can be used to describe the
entire history of the theory of the temperature correction.
However, there is no doubt that the issues brought up in \cite{1}
are important.

The purpose of our calculation is to take a different approach and
to study the low-frequency behavior of the correction in order to
understand its character. We have shown that the finite
temperature correction in \cite{1} is a low-frequency phenomenon.
The frequency is sufficiently low so that treating the plates as
bulk dielectrics is not valid.  By use of a more realistic
description of the field interaction with the plates we show that
the modes between metallic plates of finite conductivity produce a
finite temperature correction in close agreement with the
perfectly conducting case.  The principal difference between our
result and the previous work is that we have allowed for the
possibility that the derivatives of the fields at the conducting
boundary are discontinuous.  This possibility exists because the
fields produce currents in the conducting plates that are
discontinuous across the boundary between the vacuum and the
conductor.  Although it is tempting to model the finite
conductivity as a modification to the dielectric permittivity,
such a model fails when the mean free path of the conduction
electrons exceeds the penetration depth of the electromagnetic
field, and thus fails for frequencies of interest for the thermal
correction to the $TE$ electromagnetic mode.

We have shown that the conducting boundary conditions that are
applicable for frequencies where the $TE$ mode thermal correction
has its significant contribution lead to a net increase of the
$TE$ mode force, and is of the same magnitude as the perfectly
conducting case.  This result is in agreement with the
experimental results reported in \cite{2}.  However, additional
and improved experiments with large plate separations (greater
than 2 $\mu$m) with both conducting and dielectric plates would
provide the definitive test.  A particularly tempting dielectric
would be diamond which offers both theoretical and experimental
benefits.

\begin{figure*}
\includegraphics{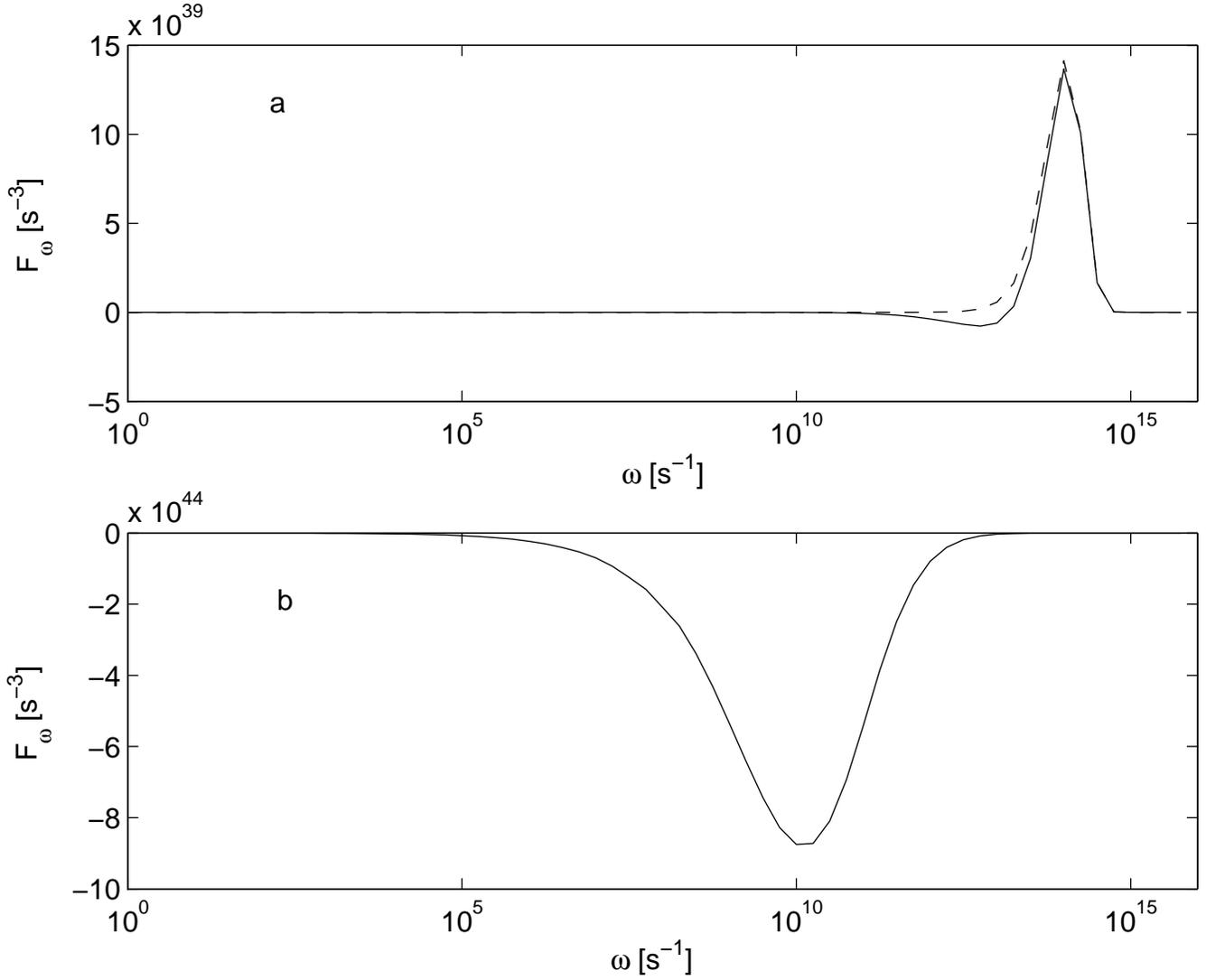}
\caption{The net finite-temperature contribution to the Casimir
force is determined by $F=(\hbar/\pi c^3)\int_0^\infty F_\omega
d\omega$ and is attractive when $F>0$. a: The two curves represent
the $C_1$ path for perfectly conducting plates (dashed curve) and
for plates with permittivity given by Eq. (\ref{epsilon})(solid
curve). The net force force for the latter is $0.95$ times the
perfectly conducting case. b: For a perfect conductor, the $C_2$
integral is zero. The net contribution from the $C_2$ path is
$-169$ times the perfectly conducting contribution from the $C_1$
path, and its addition to the $TE$ mode zero-point contribution
reduces the net $TE$ mode force to nearly zero, which is the
result obtained in \cite{1}. All are for $a=1\ \mu$m, $T=300$ K.}
\end{figure*}

\begin{figure*}
\includegraphics{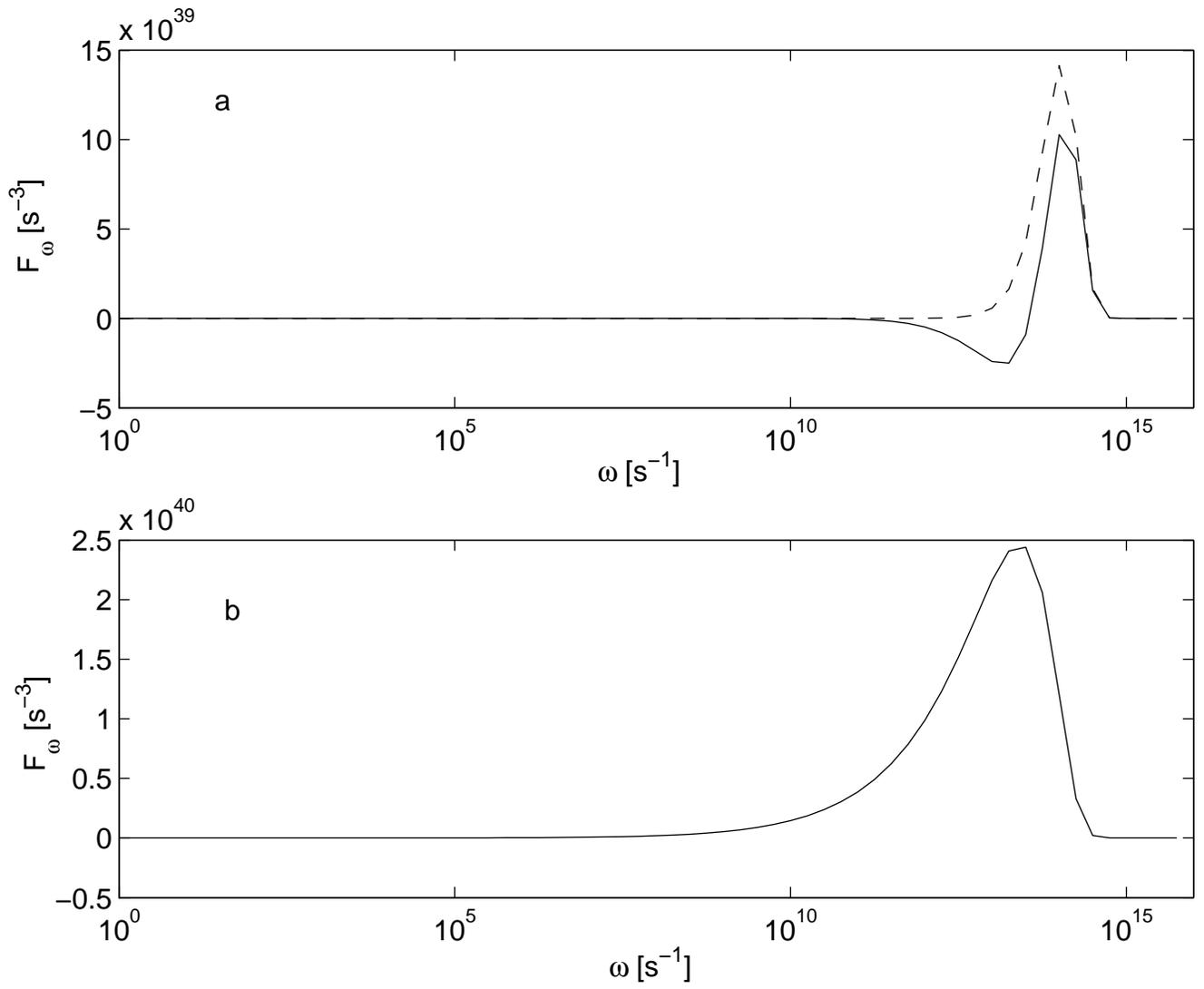}
\caption{Numerical results for $F_\omega$ using the finite
conductivity boundary conditions.  The integrated force for the
$C_2$ path contribution is $1.47$ times greater than the $C_1$
integration, and the total net force for both paths is $1.75$
times greater than the perfectly conducting case. Treatment of the
plates as conducting metals fails above $\omega =10^{14}\ {\rm
s}^{-1}$. All are for $a=1\ \mu$m, $T=300$ K.}
\end{figure*}

\end{document}